\documentclass[runningheads]{llncs}
\usepackage{graphicx}
\usepackage{array}
\usepackage{tcolorbox}
\usepackage{multirow} 
\usepackage[utf8]{inputenc}
\usepackage[colorlinks,citecolor=red,urlcolor=blue,bookmarks=false,hypertexnames=true]{hyperref}
\usepackage{booktabs,todonotes}
\usepackage[super]{nth}
\usepackage[english]{babel}
\usepackage[autostyle]{csquotes}
\usepackage{siunitx}
\usepackage{setspace}
\usepackage{academicons}
\usepackage{svg}
\usepackage{orcidlink}
\usepackage{amsmath}
\usepackage{mathtools}
\usepackage{amssymb}
\usepackage{amsfonts}
\usepackage{mathrsfs}
\usepackage{tikz}
\usepackage{pgfplots}
\usepackage{colortbl}
\usepackage[english]{babel}
\usepackage{booktabs}
\usepackage{tabularx}
\usepackage{siunitx}
\usepackage{caption}
\usepackage[super]{nth}
\usepackage[section]{placeins}
\usepackage{xspace}
\usepackage[utf8]{inputenc}
\usepackage{algorithm}
\usepackage{algpseudocode}
\usepackage[section]{placeins}
\usepackage{xcolor,pifont}
\pgfplotsset{width=5cm,compat=1.14}
\usepackage{sidecap}
\usepackage{placeins}
\usepackage{tikz}
\usetikzlibrary{decorations.pathreplacing}
\usetikzlibrary{fit}
\usetikzlibrary{shapes}
\usepackage{textcomp}
\usetikzlibrary{decorations.pathreplacing}
\usetikzlibrary{fadings}
\usepackage{mdframed}
\usepackage{mdframed}
\usepackage{color}
\usepackage{fancyvrb}
\DefineVerbatimEnvironment{HighlightCode}{Verbatim}{frame=lines,fontsize=\small}
\usepackage{listings}

\lstdefinestyle{reqStyle}{
    basicstyle=\ttfamily\footnotesize,
}

\mdfdefinestyle{reqstyle}{%
    frametitlerule=true,%
    frametitlebackgroundcolor=gray!10,
    innertopmargin=\topskip,
    frametitlefont={\itshape},
    frametitleaboveskip = 2pt,
    frametitlebelowskip = 2pt,
    innerleftmargin = 5pt,
    innerrightmargin = 5pt,
    innertopmargin=3pt,
    innerbottommargin=3pt,
    font=\sffamily\footnotesize,
    nobreak
}
\mdtheorem[style=reqstyle]{acslspecification}{ACSL specification}
\mdtheorem[style=reqstyle]{hlnlspecification}{High-Level Natural Language specification}
\mdtheorem[style=reqstyle]{llnlspecification}{Low-level Natural Language specification}

\newenvironment{boxedsection}
{\begin{mdframed}[linewidth=1pt,linecolor=black]}
{\end{mdframed}}

\begin{document}

\title{Towards Specification-Driven LLM-Based Generation of Embedded Automotive Software}



\authorrunning{M.S. Patil\and G. Ung\and M. Nyberg}

\author{Minal Suresh Patil\inst{1}\inst{2}\thanks{Work was done while the author was at Scania}, Gustav Ung\inst{2} \and Mattias Nyberg\inst{2}}

\institute{Umeå Universitet, 90187 Umeå, Sweden\\
\email{minalsp@cs.umu.se}
\and
Scania, Granparksvägen 10, 15148 Södertälje, Sweden\\
\email{\{gustav.ung, mattias.nyberg\}@scania.com}}
\maketitle              
\begin{abstract}
The paper studies how code generation by LLMs can be combined with formal verification to produce critical embedded software. 
The first contribution is a general framework, \texttt{spec2code}, in which LLMs are combined with different types of critics that produce feedback for iterative backprompting and fine-tuning. The second contribution presents a first feasibility study, where a minimalistic instantiation of \texttt{spec2code}, without iterative backprompting and fine-tuning, is empirically evaluated using three industrial case studies from the heavy vehicle manufacturer Scania. The goal is to automatically generate industrial-quality code from specifications only. Different combinations of formal ACSL specifications and natural language specifications are explored. The results indicate that formally correct code can be generated even without the application of iterative backprompting and fine-tuning.
\end{abstract}

\section{Introduction}
\label{sec:Introduction}


Recent advancements in Large Language Models (LLMs) have shown promising, and sometimes astonishing, results in 
code generation \cite{vaithilingam2022expectation,ross2023programmer}. However,
from several studies \cite{tambon2024bugs,zhong2024can}, it is also clear that it is hard to guarantee code correctness and quality. In the area of automotive embedded systems, correctness and quality of the software are crucial. To be more specific, by \emph{correctness} we here mean functional correctness with respect to functional specifications and also absence of errors that may cause safety and cybersecurity issues. By \emph{quality}, we mean all other properties typically expected in embedded code, as defined in coding standards and guidelines such as MISRA-C~\cite{MISRA2004} and \enquote{the power of 10} rules \cite{holzmann2006power}.

In the present paper, we consider the problem of using LLMs to generate source code for critical embedded  software. We make the following two contributions:
\begin{itemize}
    \item We introduce the \texttt{spec2code} framework. This framework is an adaptation of the previously presented LLM-Modulo Framework from the position paper~\cite{kambhampati2024llms} in the area of planning. In \texttt{spec2code}, LLMs are combined iteratively with \emph{critics}, such as software verifiers, to produce high-quality, correct software.  
    \item Based upon industrial case studies, we investigate the feasibility of the \texttt{spec2code} framework by considering a minimalistic instantiation of it, in which we use non-iterative prompting, exclude fine-tuning, and focus mainly on formal functional correctness of the code. 
\end{itemize}

In the second contribution, within the context of the \texttt{spec2code} framework, we explore if it is at all feasible to have LLMs generate automotive safety-critical embedded C code from specifications. 
We consider specifications being given in natural language (NL) and in the formal ANSI/ISO C Specification Language (ACSL) \cite{baudin2021acsl}. To evaluate the correctness of the code generated, we apply deductive verification and the tool Frama-C \cite{frama-c-manual} to verify whether the code implements given ACSL specifications.
Since this is our first experiments with \texttt{spec2code}, and the focus is exploring its fundamental feasibility, we have, as stated above, chosen to just implement a version without iterative backprompting. 

For the evaluation, we use three industrial case studies from the heavy vehicle manufacturer Scania. Each case study consists of a 
single software \emph{module}, i.e. a pair of a \texttt{.c}-file and a \texttt{.h}-file, from real production software. From the specifications of each such module, we try to generate functionally equivalent code by using two different LLMs, namely \textsf{gpt-3.5} and \textsf{gpt-4-turbo}.
We aim to address the following research questions:
\begin{itemize}
\item \textbf{RQ1:} How can the integration of formal verification tools, such as Frama-C, with off-the-shelf LLMs be combined to automatically generate formally verified C code from ACSL specifications, along with high- and low-level natural language software specifications?
\item \textbf{RQ2:} How can specifications be effectively translated into prompts or inputs for LLMs to ensure the generation of correct and verified C code?
\item \textbf{RQ3:} How do we assess the functional correctness and quality of the generated code?
\end{itemize}

\begin{figure}[h!]
\centering
 \includegraphics[scale=0.07]{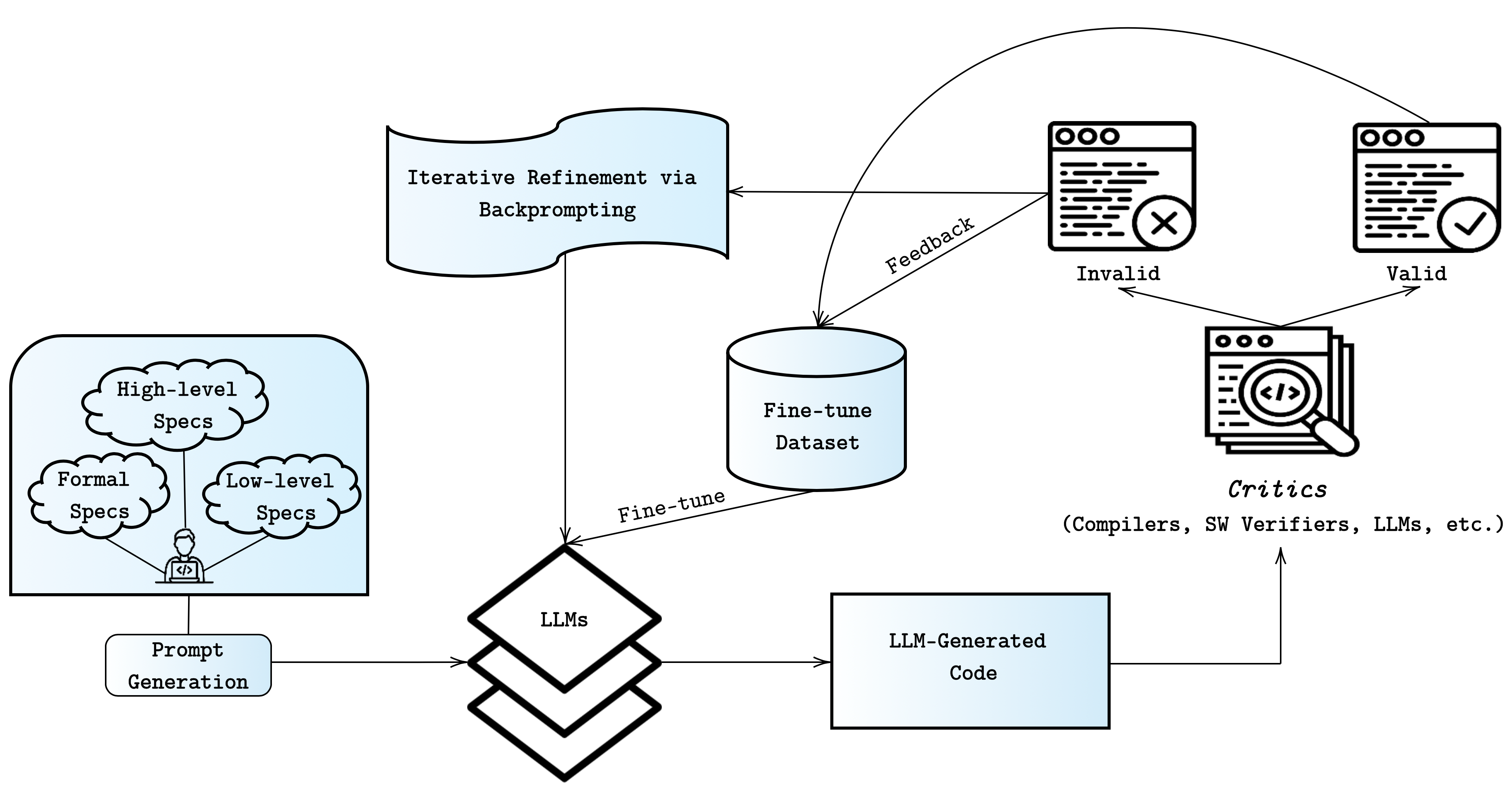}
     \caption{Overview of \texttt{spec2code} framework}
     \captionsetup{justification=centering}
\label{fig:mesh1}
\end{figure}

The \texttt{spec2code} framework is, to the best of our knowledge, the first presented general framework focusing on using LLM's to generate critical embedded software from specifications. A related work is the so-called TriCo methodology~\cite{trico} 
envisioning how LLMs and formal methods can be exploited to \emph{co-pilot} the 
simultaneous development of the three artifacts implementation, specification, and tests. There are however noticeable differences between TriCo and \texttt{spec2code}. Firstly, TriCo aims at \emph{supporting} the programmer in writing code by giving suggestions, while \texttt{spec2code} aims at \emph{replacing} the programmer with LLM- and formal-based technology, and thus, only relies on having developers creating specifications. Secondly, while TriCo emphasizes the triple including test cases, \texttt{spec2code} relies on only formal formal verification to guarantee correct code, thus eliminating the need for testing. 

The paper is organized as follows. In Section~\ref{sec:Background}, background information on deductive verification of \texttt{C} programs and LLMs are presented. Then, in Section~\ref{sec:spec2code}, we present the first contribution, i.e. the \texttt{spec2code} framework. Section~\ref{sec:caseStudies}, \ref{sec:experimental}, and \ref{sec:results} then describe the second contribution, i.e. the case study on using the minimalist instantiation of the \texttt{spec2code} framework applied to real industrial code. Finally, Section~\ref{sec:discussion} and \ref{sec:conclusions} contains a discussion and the conclusions of the paper.


\section{Background}
\label{sec:Background}
\subsection{Deductive Verification of \texttt{C} Programs}
\textsc{Frama-c} \cite{frama-c-manual} is an open-source, extensively developed framework that verifies  \texttt{C} programs annotated with ACSL~\cite{baudin2021acsl} specifications. Its \textsc{WP} plug-in enables users to verify that \texttt{C} code meets ACSL specifications through deductive verification~\cite{ahrendt2016deductive,hahnle2019deductive}. Specifically, \textsc{WP} utilizes an advanced form of weakest precondition calculus~\cite{wp-manual,leino2005efficient}, which inspired its name. ACSL specifications are placed within special comments \enquote{\texttt{/*@ \ldots*/}} and primarily include function contracts and code annotations. Function contracts feature pre-conditions (\texttt{requires} clauses) and post-conditions (\texttt{ensures} clauses), which are pure logical formulas that need to be verified before and after any function call. The \texttt{assigns} clause specifies the frame conditions of the function, i.e. the set of memory locations, including global variables and pointers, that are allowed to be modified by the function. Code annotations, such as \texttt{assert} clauses, are pure logical formulas linked to specific points in the program that must be validated on every execution path that passes through these points. \emph{Ghost variables} are variables that are evaluated in a static runtime environment disjoint from the normal heap and stack. These can be related to the normal program variables by \emph{ghost statements}, but may not affect the program. 


\subsection{Large Language Models}
In recent years, Large Language Models (LLMs) have shown impressive progress in a wide range of tasks. Notably, there has been a significant increase in the number of LLMs designed specifically for coding tasks, particularly in the area of code generation. LLMs utilize Transformers~\cite{vaswani2017attention}, which are the most advanced neural architecture in natural language processing. Additionally, Transformers have proven highly effective in addressing classic problems in verification~\cite{cosler2023iterative,hahn2020teaching}, reasoning~\cite{lewkowycz2022solving}, and in the auto-formalization of mathematics and formal specifications~\cite{hahn2022formal,cosler2023nl2spec}. OpenAI's \textsf{gpt-3.5} and \textsf{gpt-4-turbo} build on the pre-trained GPT-3 model, with additional fine-tuning using Reinforcement Learning with Human Feedback (RLHF)~\cite{ouyang2022training}. Although they are not specifically optimized for code generation, both models have shown impressive performance on various related tasks~\cite{olausson2023demystifying,schafer2023empirical}. 

\subsection{Prompting LLMs}
Prompts are user-provided inputs, such as queries and instructions that guide LLMs and instruct their behavior for specific tasks. Previous work~\cite{wei2023zero,wang2022towards} have demonstrated that the quality of input prompts plays a crucial role in the performance of LLMs, significantly influencing the quality of the output. A prompt template is often used to generate prompts in a consistent and reproducible way. Chain-of-Thought (CoT) prompting~\cite{shi2022language} has significantly enhanced zero- and few-shot performance in various complex reasoning tasks, such as arithmetic, commonsense, symbolic, and logical reasoning. A crucial feature of CoT prompting is the use of rationales, which demonstrate the step-by-step reasoning process. In this work, we design our prompts based on Zero-shot-CoT~\cite{kojima2022large}, which involves merely introducing the rationale-triggering sentence \textit{~\enquote{Let's think step by step}} to the LLMs, which leads to significant improvements in zero-shot performance.  


\section{\texttt{Spec2code} Framework}
\label{sec:spec2code}
Kambhampati et al.~\cite{kambhampati2024llms} examines the role of LLMs in planning tasks known as \textit{LLM-Modulo Framework}. It concludes that while LLMs cannot independently plan or verify, they can assist within a hybrid framework that integrates LLMs with external \emph{critics}.

Building on this core principle, we introduce \texttt{spec2code} which aims to leverage the capabilities of LLMs to generate code from both informal and formal specifications via prompting. \texttt{spec2code} aims to make use of the generative power of LLMs while ensuring the functional correctness and quality of the code generated through software verifiers. 

Figure \ref{fig:mesh1} shows the \texttt{spec2code} framework that involves a systematic method for generating code from specifications using LLMs. The process begins with the verification engineer writing both high-level and low-level specifications in natural language, as well as formal specifications in ACSL. These specifications describe the desired functionality and behavior of the code to be generated.

High-level specifications are typically abstract and outline the module's objectives and functionalities, whereas formal specifications describe the precise semantics in a formal language of software behavior. Low-level specifications describe the implementation and specific requirements the module must achieve to realize the high-level specifications. Examples of these specifications are provided in Section~\ref{sec:specificationtypes}.
 
By using a carefully crafted prompt that employs one or a combination of prompting strategies such as in-context learning, zero-shot/few-shot prompting, or Chain-of-Thought, these specifications are used to generate initial programs using LLMs. These programs are first checked for successful compilation by a compiler, and then a software verifier assesses the functional correctness of these programs. Here, we term the compiler and the software verifier as critics. If a program fails to compile successfully or fails the verification process, the feedback provided by the compiler or the software verifier is used as part of the fine-tuning dataset. This dataset is continuously updated along with the successfully verified programs to further fine-tune the LLMs. Furthermore, we employ iterative backprompting to refine our initial prompt based on feedback from various sources that may come from the compiler, counterexamples produced by software verifiers such as Frama-C, code quality assessments provided by other LLMs (which can be further optimized using prompt optimization techniques such as TextGrad and DSPy~\cite{yuksekgonul2024textgrad,khattab2023dspy}), or human evaluators. By continuously incorporating feedback from these critics, we enhance the accuracy and quality of the prompts. This refinement process ensures that each iteration of both the LLM-generated code and the prompts aligns closely with the desired code quality and standards.

Building upon this iterative refinement process, we can further fine-tune the LLMs using a supervised approach known as Supervised Fine-Tuning (SFT) and then perform preference optimization to produce preferred outputs~\cite{ouyang2022training}. This process requires data from different sources and feedback mechanisms. During the SFT stage, we collect data that includes paired datasets of specifications (both natural language and ACSL), and verified C code implementations; functional safety requirements (e.g., ISO 26262 compliance~\cite{international2018iso}); industry-standard coding guidelines (e.g., MISRA C~\cite{MISRA2004}); and successfully verified C code with common verification properties such as memory safety and absence of runtime errors. This step ensures the LLMs generate appropriate programs for given inputs. Feedback data will consist of failed verification attempts with counterexamples along with explanations of why the verification failed, human expert reviews or comments via iterative backprompting, compiler warnings and errors, and any feedback from other LLMs. We then construct preference pairs, i.e., preferred (verified, correct) vs. non-preferred (failed verification, incorrect) programs, which are optimized to produce preferred outputs using optimization techniques such as Direct Preference Optimization (DPO)\cite{rafailov2024direct}. This approach aligns the LLM to produce verified code by leveraging multiple feedback sources and domain knowledge. This process necessitates a robust pipeline for code generation, verification, feedback collection, and model updating, with careful balancing of feedback sources to prevent overfitting.

The main objective of the \texttt{spec2code} framework is to produce high-quality, specification-compliant code through a combination of initial LLM-based code generation from specifications and through iterative refinement via backprompting. Our framework integrates high-level, low-level and formal specifications with LLM-generated code, iterative refinement via backprompting, and a fine-tuning dataset. The iterative nature and continuous improvement of the fine-tuning dataset ensure that the generated code progressively aligns more closely with the intended specifications, resulting in reliable, safe, and trustworthy code. 


\section{Case Studies}
\label{sec:caseStudies}

Our case study focuses on three automotive control modules obtained from Scania: the Steering Fluid Level Detection (SFLD) module, the Brake Light Activation (BRAK) module, and the Power Steering Backup (STEE) module. We assess the correctness and quality of the generated code. This work not only evaluate the feasibility of using LLMs for such tasks but also aims to identify potential limitations and areas for improvement in the context of automotive software engineering. 

\subsection{Code Style of Application Modules in Embedded/Automotive Software}
\begin{figure}
    \centering
    \begin{tikzpicture}
    \node[draw, rectangle,minimum height=2em] (sche) {Scheduler};
    \node[below = 3em of sche, draw, ellipse,outer sep =1.5pt] (appl) {appl};
    \node (diag) [right = 2em of appl, draw, rounded corners,outer sep=1.5pt] {Diagnostics};
        \node (rtdb) [above = 1em of diag, draw, rounded corners,outer sep=1.5pt] {RTDB};
    \node (llsw) [above = 1.5em of rtdb, draw, rounded corners,align=center,outer sep=1.5pt] {Low-level\\SW/HW};
    \node (can) [right = 3em of llsw, draw, rounded corners,outer sep=1.5pt] {CAN};
    \node (sens) [below = 1em of can, draw, rounded corners] {Sensors};
    \node (actu) [below = 1em of sens, draw, rounded corners,outer sep=1.5pt] {Actuators};
    \draw[->, dashed,line width=0.4mm] (sche) to  node[left,align=center] (ten) {call main\\[-2pt]every 10\,ms} (appl);
    \draw[<->] (llsw) to node[above,xshift=2pt] {r/w} (can);
    \draw[<-] (llsw) to node[above,yshift=-2pt,xshift=1pt] {r} (sens);
    \draw[->] (llsw) to node[below] {w} (actu);
    \draw[<->] (rtdb) to node[left] {r/w} (llsw);
    \draw[<->] (appl) to node[above] {r/w} (rtdb);
    \draw[<->] (appl) to node[below,yshift=1pt] {r/w} (diag);
    \node[fit={([shift={(-10.3mm,0mm)}]sche.west) (appl) (llsw) (diag)},draw=gray!90] (ecubox) {};
    \node[below right=0.022em and 0.022em, inner sep=4pt, fill=gray!20] at (ecubox.north west) {ECU};
\end{tikzpicture}
    \caption{ECU System architecture}
    \label{fig:appl-sys-arch}
\end{figure}
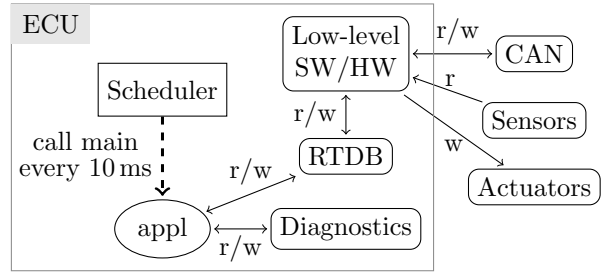
All case study modules in this study are written in C~\cite{ISO:C99}, and are developed according to internal coding
guidelines largely identical to those of MISRA-C~\cite{MISRA2004}. The modules are executed in two of several Electronic Control Units (ECUs) in
the embedded system of Scania trucks. Each ECU has a static scheduler which repeatedly calls the main entry point of each application module once every 10 ms. The control flow is simple, containing no recursion or loops. The module (in isolation) is strictly sequential, although as part of the larger system it is not.
Other than the scheduler, the application modules communicates directly
with other application modules and infrastructure software. The two main mechanism for internal communication within the ECU are a diagnostics module, and a real-time database (RTDB) module which facilitates communication between the different application level modules as shown in Figure~\ref{fig:appl-sys-arch}. The external communication between ECUs is done by CAN~\cite{ISO:CAN}, and the application can indirectly interact with the physical environment by sensors and actuators. The implementation is generally structured as follows: when the main function is called, all inputs from the system are read by calling functions in RTDB, then several computations are made using these inputs and static variables within the module, before finally writing all outputs to RTDB. Reads and writes to the diagnostics module occur throughout computing results, and may be part of, or the result of, those computations.

\subsection{Working with the case studies}
All three case studies are taken from Scania's development repository. Each case study has a set of natural language specifications of varying quality associated with it. From these natural language specifications, the ACSL specifications were derived by hand.

We restrict all case studies to one single translation unit each, meaning that the interaction with library code has been abstracted away. For example, the interaction with RTDB has been removed, and instead the relevant signals are assumed to be globally defined static variables. Importantly, the actual code functionality has not been changed, and therefore the modified code follows the original code.


\subsection{Case study 1: Oil Level Warning (SFLD)}
The oil level warning module is a software responsible for emitting diagnostics warnings whenever the oil level has been low for a specific amount of time. The original module has ~200 LoC, not including declarations and type definitions from header files. There are 1 high-level natural language specification, 11 low-level natural language specifications, and 10 derived ACSL specifications for this module.
\subsection{Case study 2: Brake Light Activation (BRAK)}
The brake light module is a software responsible for activation of brake lights. The original module has ~400 LoC not including declarations and type definitions from header files. In total, there are 1 high-level natural language specification, 17 low-level natural language specifications, and 17 derived ACSL specifications for this module
\subsection{Case study 3: Power Steering Backup (STEE)}
The power steering backup module is software responsible for activation of a backup system in case the primary power steering system fails. This module and its specifications has been thoroughly studied in a previous paper~\cite{steepaper}. In the present paper, a modified version has been used. There are 1 high-level natural language specification, 5 low-level natural language specifications, and 5 derived ACSL specifications for this module. This version of the module has ~150 LoC.

\section{Experimental Setup}

\begin{figure}[h!]
\centering
 \includegraphics[scale=0.08]{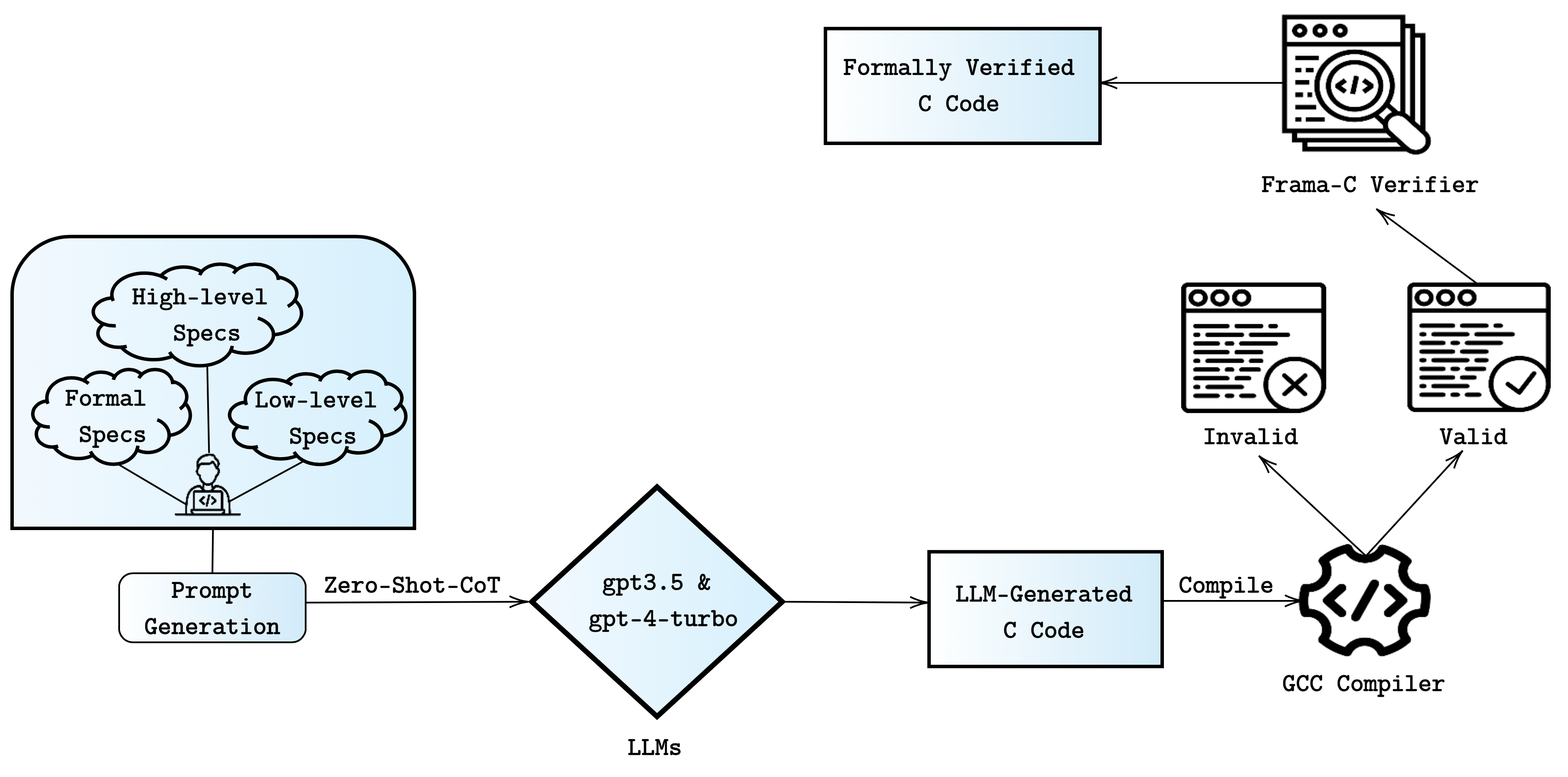}
     \caption{Overview of the minimalist version of \texttt{spec2code} framework}
     \captionsetup{justification=centering}
\label{fig:mini}
\end{figure}

\label{sec:experimental}

As shown in Figure~\ref{fig:mini}, we evaluate a minimalist instantiation of \texttt{spec2code} without iterative prompting and fine-tuning. With a minimalistic approach, we aim to explore how the latest State-of-the-art (SotA) Large Language Models (LLMs) perform in a real-world industrial setting. This particular instantiation includes GCC as the C compiler and Frama-C as the software verifier, which serve as the critics. For our evaluation, we also use the \emph{diffkemp}~\cite{diffkemp} program equivalence tool, and human code quality inspection. 

\subsection{Specification Types} 
\label{sec:specificationtypes}
\begin{enumerate}
    \item \texttt{High-Level Natural Language (HLNL)}: Set of high-level specifications that captures the overall functionality and constraints of the system focusing on what the system needs to achieve. A high-level specification is shown below.
    \begin{hlnlspecification}\label{hlnl:example1}
        If there is a brake light request, then the truck lights shall be activated.
    \end{hlnlspecification}
    \item \texttt{Low-Level Natural Language(LLNL)}: Set of low-level specifications which combined aims to realize the high-level specifications. A low-level specification related to the high-level specification above is shown below.
    \begin{llnlspecification}\label{llnl:example1}
        If the operational state is normal or emergency with limited functionality, and the supply voltage is not low, then the brake lights should be enabled.
    \end{llnlspecification}
    \item \texttt{ACSL}: Set of formal specifications in written in ACSL. An ACSL specification derived from the low-level specification above is shown below.
    \begin{acslspecification}\label{acsl:example1}
        \begin{lstlisting}[style=reqStyle,belowskip=0pt,aboveskip=0pt]
((gh_operationalState == NORMAL_OPERATION 
|| gh_operationalState == EMERGENCY_STOP_LIMITED)
&& gh_supplyVoltageLevel != VOLTAGE_LOW)
==> gh_brakeLightEnabled == TRUE;  
        \end{lstlisting}

    \end{acslspecification}
\end{enumerate}
In our experiments, we use all possible combinations of specifications to thoroughly evaluate the LLM's capability to generate functionally correct code with respect to the specifications. By incorporating different specification types, we aim to evaluate the effectiveness in producing functionally correct code across various specification categories.

\subsection{LLM Selection}
In our experiments, we evaluate our framework on current SotA LLMs namely \textsf{gpt-4-turbo}~\cite{openai_gpt4-turbo} and \textsf{gpt-3.5}~\cite{openai_gpt3_5}. These models were chosen because they offer API access, which allows us to leverage their capabilities without the need for high memory requirements. 
For \textsf{gpt-4-turbo} and \textsf{gpt-3.5} we configure the parameters of the LLMs as follows: $max\_token: 4096, temperature: 0.8, presence\_penalty: 0.5, frequency\_penalty: 1$, and $top\_p: 1.$ A temperature of $0.8$ is preferred for our experiments because it introduces controlled randomness, allowing for more diverse and creative programs while still maintaining coherence.

\subsection{Prompt Design}
We designed a prompt based on the Zero-shot-CoT prompting technique to guide the LLM in generating code from ACSL specifications, as well as high-level and low-level specifications provided in natural language.
We first specify the \textit{system prompt} to define the intended role, capabilities, and behavior of the LLM. Next, we specify the \textit{user prompt} that instructs the LLM to complete the given task, i.e., completing the function definition from specifications. We provided the structure of the C program, including the header, global variables, and function header, as part of the prompt (See \hyperref[subsec:appendix]{\nameref*{subsec:appendix}} for prompt template).

\subsection{Evaluation of Generated Code} 
In all case studies, each combination of model and specification produces a generated program which is then compiled. If the compilation is successful, then the program is evaluated in three different ways: functional correctness, program equivalence, and code quality. \\ \\
\textbf{Functional correctness}. The functional correctness of the program is verified with respect to its ACSL specification by using the weakest-preconditions calculus plugin (WP) in Frama-C. We choose Frama-C as a verification framework because we deem it to be mature and of state-of-the-art. In our evaluation we use the Frama-C version 27.1 with the backend solvers Z3~\cite{de2008z3} and Alt-Ergo~\cite{conchon2018alt}. The results can be seen in section~\ref{sec:results}. 
\\ \\
\textbf{Program equivalence.}
The functional correctness verification shows whether the implementation satisfies the specification, but it is possible that the specification itself might be incomplete. By incompleteness, we mean that the set of program behaviors specified by the specification might not cover the intended program behaviors of the code.We evaluate the \textit{relative correctness} of the original code from the Scania repository and the code generated by the LLM using program equivalence based on program semantics.

All case studies include one handwritten program each, which is verified w.r.t. its formal specification, and is considered to be the reference program. The generated programs are compared to the relevant reference program, and any semantic deviation is captured. Concretely, the evaluation is done by using the tool \emph{diffkemp}~\cite{diffkemp}. This tool can show program equivalence on instruction-level, for programs written in LLVM byte-code. A set of semantics preserving transformations is used to accomplish this task. Hence, the tool can fail to show equivalence either because the programs are not equivalent, or because the programs are too different structurally. 
\\ \\
\textbf{Code Quality.}
The code quality is evaluated quantitatively by counting the lines of code (LoC), and qualitatively by evaluating its conformance with \enquote{the power of 10} rules \cite{holzmann2006power}.  Only the lines of code in the implemented function are counted, and all empty lines and comments are removed. Generally a longer program implementing the same functionality as a short program can be considered to be more complex, and of lower quality. However, analysis of LoC is surely not sufficient for evaluating the code quality. Hence, we also manually inspect the generated code and evaluate its conformance with \enquote{the power of 10} rules. A summary of the rules is the following:
\begin{enumerate}
    \item Avoid complex flow constructs, such as goto and recursion.
    \item All loops must have fixed bounds. 
    \item Do not use dynamic memory allocation after initialization.
    \item Restrict size of function to around 60 LOC.
    \item (Use a minimum of two runtime assertions per function.) \textbf{Not used.}
\item Restrict the scope of data to the smallest possible.
\item Check the return value of all non-void functions, or cast to void to indicate the return value is useless.
\item Use the preprocessor sparingly.
\item Limit pointer use to a single dereference, and do not use function pointers.
\item Compile with all possible warnings active; all warnings should then be addressed before release of the software. 
\end{enumerate}
As indicated in the list above, we do not use Rule 5. We argue that the use of this rule is questionable, at least in the context of automotive embedded software. In C/C++, it is in fact generally recommended to not use assertions in production code, e.g. see \cite{enwiki:1179241560}. Rule 8 is deemed to be followed when each instance of preprocessor usage can be justified, and is judged to be simpler than an alternative approach. If a file is not compilable, then no code quality analysis is performed.

\section{Results}

In this section, we discuss the results of our research by placing them in the context of the three research questions:

\begin{enumerate}
    \item Capability of current SotA LLMs to generate code based on formal, high-level, and low-level specifications \textbf{(RQ1)}.
    \item Effectiveness of input prompt design on the LLM's performance \textbf{(RQ2)}.
    \item Evaluation of both functional correctness and overall quality of the LLM-generated code \textbf{(RQ3)}.
\end{enumerate}

\noindent It is worth noting that all compilable programs produced by the LLM adhered to the \enquote{power of 10} rules, and no warnings were issued by the GCC compiler. Consequently, our analysis tables focus solely on the Lines of Code (LoC) metric.

\label{sec:results}
\FloatBarrier
\begin{table}[h]
\centering
\small
\begin{tabular}{|>{\centering\arraybackslash}m{2.15cm}|>{\centering\arraybackslash}m{2.15cm}|>{\centering\arraybackslash}m{1.5cm}|>{\centering\arraybackslash}m{1.5cm}|>{\centering\arraybackslash}m{1.5cm}|>{\centering\arraybackslash}m{1.5cm}|}
\toprule 
\texttt{Model} & \texttt{Specification Type} & \texttt{Compiled} & \texttt{Eq. Check} & \texttt{Verified (Proved Goals)} & \texttt{LoC} \\ 
\midrule
\midrule
\texttt{GPT-4-turbo} & \texttt{ACSL}  & Yes & Not Eq  &  \cellcolor{green!55}33 / 33  & 35 \\
& \texttt{HLNL}   & Yes  & Not Eq  & 5 / 23 & 25 \\
& \texttt{LLNL}  & Yes  & Not Eq & 23 / 29  & 54 \\
& \texttt{ACSL + HLNL + LLNL}  & Yes & Not Eq & 23 / 29  & 35 \\
& \texttt{HLNL + LLNL}   & Yes  & Not Eq  & 23 / 29 & 40 \\   
& \texttt{ACSL + LLNL}   & Yes & Not Eq  & 18 / 25  & 36 \\   
& \texttt{ACSL + HLNL}   & Yes  & Not Eq  & 24 / 29 & 37 \\  

\midrule

\texttt{GPT-3.5} & \texttt{ACSL}  & Yes & Not Eq & 21 / 27 & 37 \\
& \texttt{HLNL}   & Yes   & Not Eq & 11 / 22  & 25 \\
& \texttt{LLNL}  & No  & n/a & n/a & n/a \\
& \texttt{ACSL + HLNL + LLNL}  & No  & n/a & n/a & n/a \\
& \texttt{HLNL + LLNL}   & No  & n/a & n/a & n/a \\   
& \texttt{ACSL + LLNL}   & No  & n/a & n/a & n/a \\   
& \texttt{ACSL + HLNL}   & Yes & Not Eq & 18 / 27 & 35 \\  





\bottomrule
\end{tabular}
\caption{Steering Fluid Level Detection (SFLD) module}
\label{tab:sfld}
\end{table}
\FloatBarrier

\FloatBarrier
\begin{table}[ht]
\centering
\small
\begin{tabular}{|>{\centering\arraybackslash}m{2.15cm}|>{\centering\arraybackslash}m{2.15cm}|>{\centering\arraybackslash}m{1.5cm}|>{\centering\arraybackslash}m{1.5cm}|>{\centering\arraybackslash}m{1.5cm}|>{\centering\arraybackslash}m{1.5cm}|}
\toprule 
\texttt{Model} & \texttt{Specification Type} & \texttt{Compile} & \texttt{Eq. Check} & \texttt{Verified (Proved Goals)} & \texttt{LoC} \\ 
\midrule
\midrule

\texttt{GPT-4-turbo} & \texttt{ACSL}  & Yes  & Not Eq  & 41 / 50 & 48 \\
& \texttt{HLNL}   & Yes  & Not Eq  &  13 / 26  & 21  \\
& \texttt{LLNL}  & Yes  & Not Eq & 39 / 40 & 48 \\
& \texttt{ACSL + HLNL + LLNL}  & Yes  & Not Eq & 40 / 52 & 64 \\
& \texttt{HLNL + LLNL}   & Yes  & Not Eq & \cellcolor{green!55}30 / 30 & 32 \\   
& \texttt{ACSL + LLNL}   & Yes  & Not Eq & 39 / 40 & 65 \\  
& \texttt{ACSL + HLNL}   & Yes  & Not Eq & 41 / 52 & 30 \\  

\midrule

\texttt{GPT-3.5} & \texttt{ACSL}  & Yes  & Not Eq & 41 / 43 & 36 \\
& \texttt{HLNL}   & Yes  & Not Eq & 11 / 24 & 20 \\
& \texttt{LLNL}  & No  & n/a & n/a & n/a \\
& \texttt{ACSL + HLNL + LLNL}  & Yes  & Not Eq  & 39 / 40  & 44 \\
& \texttt{HLNL + LLNL}   & No  & n/a  & n/a & n/a \\   
& \texttt{ACSL + LLNL}   & Yes  & Not Eq  &  13 / 30  & 35 \\   
& \texttt{ACSL + HLNL}   & Yes  & Not Eq  &  38 / 42 & 33 \\  





\bottomrule
\end{tabular}
\caption{Brake Light Activation (BRAK) module}
\label{tab:brak}
\end{table}
\FloatBarrier

\FloatBarrier
\begin{table}[ht!]
\centering
\small
\begin{tabular}{|>{\centering\arraybackslash}m{2.15cm}|>{\centering\arraybackslash}m{2.15cm}|>{\centering\arraybackslash}m{1.5cm}|>{\centering\arraybackslash}m{1.5cm}|>{\centering\arraybackslash}m{1.5cm}|>{\centering\arraybackslash}m{1.5cm}|}
\toprule 
\texttt{Model} & \texttt{Specification Type} & \texttt{Compile} & \texttt{Eq. Check} & \texttt{Verified (Proved Goals)} & \texttt{LoC} \\ 

\midrule
\midrule

\texttt{GPT-4-turbo} & \texttt{ACSL}  & Yes  & Not Eq & 6 / 18  & 27 \\
& \texttt{HLNL}   & Yes & Not Eq  & 3 / 8 & 6  \\
& \texttt{LLNL}  & Yes  & Not Eq & \cellcolor{green!55}8 / 8 & 22 \\
& \texttt{ACSL + HLNL + LLNL}  & Yes  & Not Eq  & \cellcolor{green!55}8 / 8 & 26 \\
& \texttt{HLNL + LLNL}   & Yes  & Not Eq  & \cellcolor{green!55}8 / 8 & 24 \\   
& \texttt{ACSL + LLNL}   & Yes   & Not Eq  & \cellcolor{green!55}8 / 8 & 38 \\   
& \texttt{ACSL + HLNL}   & Yes  & Not Eq & \cellcolor{green!55}8 / 8 & 36 \\  

\midrule

\texttt{GPT-3.5} & \texttt{ACSL}  & Yes & Not Eq  & \cellcolor{green!55}8 / 8 & 17 \\
& \texttt{HLNL}   & Yes  & Not Eq & 3 / 8 & 9 \\
& \texttt{LLNL}  & Yes  & Not Eq & \cellcolor{green!55}8 / 8  & 16 \\
& \texttt{ACSL + HLNL + LLNL}  & Yes  & Not Eq & \cellcolor{green!55}8 / 8 & 16 \\
& \texttt{HLNL + LLNL}   & No  & n/a & n/a & n/a \\   
& \texttt{ACSL + LLNL}   & Yes  & Not Eq & \cellcolor{green!55}8 / 8  & 16 \\    
& \texttt{ACSL + HLNL}   &  Yes & Not Eq & \cellcolor{green!55}8 / 8 & 16  \\    





\bottomrule
\end{tabular}
\caption{Power Steering Backup (STEE) module}
\label{tab:stee}
\end{table}
\FloatBarrier

\section{Discussion}
\label{sec:discussion}
\subsubsection{LLM comparison.}
To ensure a fair comparison between \textsf{gpt-3.5} and \textsf{gpt-4-turbo}, we evaluate and report only the initial code generation response, known as $pass@1$~\cite{chen2021evaluating}. The $pass@1$ metric offers a direct measure of an LLM's capability to produce correct code on the first attempt, i.e., producing syntactically correct code that is also successfully verified by a software verifier. By focusing on the first generated code output, we highlight the LLM's efficiency and reliability in producing correct solutions without requiring further iterations thus providing an unbiased assessment of each LLM's performance under the same parameter settings, making it easier to compare their effectiveness. However, it worth noting that the $pass@1$ is a strict metric because it does not account for the LLM's performance on subsequent attempts or its ability to learn from feedback. 

In Tables \ref{tab:sfld},~\ref{tab:brak} and \ref{tab:stee} we compare the performance of \textsf{gpt-4-turbo} and \textsf{gpt-3.5} models across various specification types (ACSL, HLNL, LLNL, and their combinations) for each of the case study. 

In Table \ref{tab:sfld}, \textsf{gpt-4-turbo} successfully produces code that compiles for all specification types, while \textsf{gpt-3.5} is unable to do so with several. In terms of verification, \textsf{gpt-4-turbo} produces code that is successfully verified with ACSL specifications where Frama-C proves all 33 goals, whereas \textsf{gpt-3.5} performs less consistently, with fewer proved goals and failing to compile for multiple specification types.

In Table \ref{tab:brak}, \textsf{gpt-4-turbo} successfully compiles all specification types, while \textsf{gpt-3.5} fails to compile LLNL and HLNL + LLNL specification types. \textsf{gpt-4-turbo} generally achieves higher numbers of verified (proved) goals compared to \textsf{gpt-3.5}, with the HLNL + LLNL combination for \textsf{gpt-4-turbo} achieving 30/30 proved goals.

In Table \ref{tab:stee}, \textsf{gpt-4-turbo} successfully compiles all specification types, while \textsf{gpt-3.5} fails only for the HLNL + LLNL combination. In terms of verified proved goals, \textsf{gpt-4-turbo} successfully verifies most combined specifications except for ACSL and HLNL specification types, whereas \textsf{gpt-3.5} fails to compile the HLNL + LLNL specification type but is able to verify all specification types except the HLNL specification type.

From the results, we see that relying solely on high-level natural language specifications often leads to suboptimal results. This is likely because high-level specifications are inherently ambiguous and lack technical specificity, causing LLMs to misinterpret the specifications. Consequently, the generated code may not align with the user's actual intent.


\subsubsection{Ghost variables.}
Ghost variables in ACSL play an important role in formal verification by serving as auxiliary variables to help prove program properties. However, we observe that their use can significantly impact the verification process in code generation by LLMs since they are highly context-dependent and if the LLM misinterprets the role of ghost variables or if they are not properly specified in the prompt, the LLM tends to generate code that misuses or incorrectly implements these variables, leading to failed verification instances. Furthermore, the additional overhead or added complexity of specifications involving ghost variables can inhibit the capability of the LLMs to produce code, especially if they have not been adequately trained on similar examples. 

Thus, ensuring consistency and correct usage of ghost variables in both the specification and the generated code is crucial for successful verification of the code. Due to the mentioned challenges in using ghost variables for code generation, in all our case studies, we converted the ghost variables in the specification to concrete variables in order to ensure the LLMs produce at least that code that is able to compile successfully. 






\subsubsection{Equivalence checking.}
The program equivalence checking tool was not successful in showing equivalence in any case. By doing manual inspection, we found that this is due to several reasons. First, we must recognize that the LLM was tasked with generation of code according to the specification, and nothing more. However, while the specification is supposed to be complete for the generated program (by construction), this specification does not necessarily need to specify the complete behavior of the reference program. Therefore, the specified behavior is equivalent for both programs, but the unspecified behavior is not.



Another reason for the program inequality is due to ghost variables in the specification. As previously stated, all generated programs treat ghost variables as concrete program variables. In the reference programs, the corresponding behavior is instead implemented in terms of local variables which are not visible in the function contract due to scoping rules. Therefore, ghost annotations are used in the implementation, relating the local variables with the ghost variables. Consequently, the semantic equivalence can not be shown, because the generated program additionally affects the concrete (formerly ghost) variables, i.e. the set of concrete global variables involved in side effects is larger for the generated program.

While the results show that equivalence could not be shown between the reference program and the generated programs, equivalence could be shown between generated programs in some cases. In particular, five of the \textsf{gpt-4-turbo} generated programs for the STEE case study, all of which could be functionally verified, could be divided into three different equivalence classes. These variations clearly illustrate that multiple interpretations were allowed by the specification.



\subsubsection{Code quality}
As stated in the results, all code modules adhere to \enquote{the power of 10} rules. Despite having temporal specifications, i.e. specifications that must be interpreted over time and multiple executions of the code modules, the generated code did not contain any use of loops or recursion, in adherence to rule 1. This was probably prevented by giving the LLM context about the function execution environment such as details about the scheduler. Importantly, the generated code did not feature any dynamic memory allocations or pointers, in adherence to rules 3 and 9. It is possible that pointers were avoided due to the usage of a single function for each module, this removes the need for transferring struct data between functions, which is a normal use-case for pointers. The LLM was provided with the module interface, and the intended execution context, namely automotive software, and this is likely to have influenced the style of the generated code.

While the general style of code produced is deemed suitable for usage in automotive embedded devices, the code style was inconsistent in some cases. For example, in some cases input validation was done by using normal if-statements mixed with ternary if-statements. It is possible that the usage of ternary-statements were influenced by preprocessor macros that are defined in terms of ternary-statements.

\subsection{Threats to Validity}
\subsubsection{Internal Validity.}
A potential threat to the internal validity of our method is the natural non-deterministic behavior of LLMs, which leads to unpredictable and non-repeatable code generation. We experimented with various temperature settings and ultimately set the LLM's temperature to 0.8 to control the randomness in the code produced. Another, a potential threat to the internal validity, when evaluating performance of LLMs for code generation, is that the code examples used in the evaluation have been part of the training data. This threat is completely avoided in the experiments of the present paper since all case studies use proprietary, not published, code from the manufacturer Scania.

\subsubsection{External Validity.}
While it is technically feasible to substitute \textsf{gpt-4-turbo} and \textsf{gpt-3.5} with other LLMs used in our experiments, this may yield different results, indicating that our findings might not apply universally. It is essential to consider whether our findings are valid for other codebases at Scania and in different companies. The variability in contexts, coding standards, and organizational requirements could affect the applicability of our findings, thus necessitating further validation to ensure broader relevance.



\section{Conclusion}\label{sec:conclusions}
The paper has studied how code generation by LLMs can be combined with formal verification with the aim of producing critical embedded software.
The first contribution is a general framework \texttt{spec2code}, where LLMs are combined with different types of critics.
In the second contribution, as a first feasibility study,
a minimalistic instantiation of \texttt{spec2code}, without iterative backprompting and fine-tuning, was applied to three industrial case studies from the heavy vehicle manufacturer Scania. The goal was to, from specifications only, automatically generate industrial quality code that could be hypothetically integrated in the Scania production code.

Despite not using iterative backprompting and any fine-tuning, we were able to produce successfully compiled code in all three case studies. The code was also formally verified, for some combination of specifications, in two of the three case studies. We view this result as quite promising, although the case studies were all relatively small and of low complexity. However, we expect that with the addition of iterative backprompting and fine-tuning, which we plan as future work, the approach will also be much more powerful and become capable of handling more challenging coding problems. Additionally, we intend to include the $pass@k$ in future work as this will offer a more comprehensive understanding of the LLMs ability to generate code by accounting for the presence of a correct program among the top $k$ candidate programs. Furthermore, while we acknowledge that closed-source models for scientific evaluation may be sub-optimal, we aim to address this concern in the future work by evaluating our framework using open-source LLMs.

An important conclusion of the work is that we note that, while a human programmer utilizes specifications \emph{and} context knowledge, the LLMs in \texttt{spec2code} only have access to specifications. Thus, the correctness of the generated code is with respect to the specifications only. This implies that to have \emph{complete} specifications becomes crucial. This we judge to be rare in real industrial software development. This raises the question if the extra efforts needed to make specifications complete, is more or less compared to the effort saved by delegating the programming to LLMs. This is also something we plan to investigate in future work.



\newpage
\appendix
\begin{boxedsection}
\section*{Appendix A}\label{subsec:appendix}
\subsection*{Prompt Template for HLNL for Specification Type}
\textbf{System Prompt:}
You are an experienced verification engineer with expertise in safe embedded C programming and writing ANSI/ISO C Specification Language (ACSL) specifications for safety-critical systems. Your role is to analyze C programs with accompanying ACSL and natural language specifications. Produce complete C functions that satisfy the given specifications by following these guidelines:
\begin{itemize}
    \item Do not alter the provided specifications.
    \item Do not explain or comment on the code you produce.
    \item Do not modify any provided header files.
\end{itemize}
When given a task, focus only on implementing the required C function that meet the specifications. Prioritize safety and correctness in your implementations, ensuring that your code not only meets the given specifications but also adheres to best practices for safety-critical systems by following the 10 Rules for Developing Safety-Critical Code.

\begin{itemize}
    \item Avoid complex flow constructs, such as goto and recursion.
    \item All loops must have fixed bounds. 
    \item Do not use dynamic memory allocation after initialization.
    \item Restrict size of function to around 60 LOC.
    \item (Use a minimum of two runtime assertions per function.)
    \item Restrict the scope of data to the smallest possible.
    \item Check the return value of all non-void functions, or cast to void to indicate the return value is useless.
    \item Use the preprocessor sparingly.
    \item Limit pointer use to a single dereference, and do not use function pointers.
    \item Compile with all possible warnings active; all warnings should then be addressed before release of the software. 
\end{itemize}

\noindent \textbf{User Prompt:}
Generate the C code for a function that implements the following high-level specification.
\begin{itemize}
    \item If there is a brake light request, then the truck and trailer lights shall be activated.
\end{itemize}
\footnotesize

\begin{Verbatim}

//Header
typedef unsigned char       tB;
...

typedef struct {
    tB val;
    tU08 ss_U08;
} tBS;
...

#define TRUE                    1
#define FALSE                   0
...

/*@ 
    assigns *outSig;
    ensures \valid(outSig);
    ...
*/
extern void validateInputBool(const tBS * const inSig,
                              tBS * const outSig,
                              tB defaultNotExist,
                              tB defaultNotGood);


//Input variables
static tU08S rtdb_state;
static tBS rtdb_req;
...

//Output variables
static tBS rtdb_truck;
...

//Concrete variables
tU08 gh_operationalState;
tU08 gh_supplyVoltageLevel;
tU08 gh_brakeLightEnabled;
...

//Function
void Brak_10ms(void);

\end{Verbatim}
\end{boxedsection}

\end{document}